\documentclass[11pt]{article}

\usepackage{cite,hyperref,amsfonts,amsmath,amssymb,bm,graphicx,subfigure,wasysym,a4wide}

\begin{document}

\title{\textbf{Strong magnetic fields in compact stars as a macroscopic parity violation phenomenon}}

\author{Maxim Dvornikov$^{a,b}$\thanks{maxdvo@izmiran.ru}
\\
$^{a}$\small{\ Pushkov Institute of Terrestrial Magnetism, Ionosphere} \\
\small{and Radiowave Propagation (IZMIRAN),} \\
\small{108840 Moscow, Troitsk, Russia;} \\
$^{b}$\small{\ Physics Faculty, National Research Tomsk State University,} \\
\small{36 Lenin Avenue, 634050 Tomsk, Russia}}

\date{}

\maketitle

\begin{abstract}
We review the results in our recent works on the generation of strong magnetic fields in compact stars driven by the parity violating electroweak interactions between background fermions. The cases of a neutron star, a hybrid star, and a quark star are considered. We formulate the system of kinetic equations for the description of the spectra of the magnetic energy and the magnetic helicity, as well as for the chiral imbalances. Turbulence effects, which can be important for the evolution of small-scale magnetic field, are also taken into account. We find the numerical solution of these equations in case of large- and small-scale magnetic fields in quark matter. The applications of the obtained results for the generation of large-scale magnetic fields in magnetars and for the explanation of magnetar bursts are discussed.
\end{abstract}

\maketitle


In the wake of the discovery of anomalous X-ray pulsars (AXP) and soft gamma ray repeaters (SGR), it was commonly accepted that strong magnetic fields are important for the energy radiation by compact stars. From the point of view of modern astrophysics~\cite{Mer15} AXP and SGR are assumed to be strongly magnetized, with $B > 10^{15}\,\text{G}$, compact stars, or magnetars.
Since the discovery of magnetars, the origin of their strong magnetic fields is still a puzzle for modern physics and astrophysics~\cite{Mer15}. Indeed one should explain the enhancement of the magnetic field of a pulsar $B_0 = 10^{12}\,\text{G}$ by at least three orders of magnitude to achieve the values ​​observed in magnetars $B > 10^{15}\,\text{G}$.

There are numerous models for the generation of strong magnetic fields in magnetars. Many of these models are reviewed in Ref.~\cite{Mer15}. Recently there were several attempts to explain the generation of magnetic fields in magnetars using the elementary particle physics methods based on the chiral magnetic effect (CME)~\cite{MirSho15}. Among them one should mention Refs.~\cite{ChaZhi10,SigLei16,Yam16a}. Another approach for the generation of magnetic fields in compact stars, which is based on the field instability, driven by the parity violating interaction between background fermions, was proposed in Ref.~\cite{BoyRucSha12}.

In Refs.~\cite{DvoSem15a,DvoSem15b,DvoSem15c,Dvo16a,Dvo17}, we used the idea of Ref.~\cite{BoyRucSha12} to elaborate the model for the generation of strong large-scale magnetic fields in magnetars driven by the electroweak electron-nucleon interaction in the neutron star matter. Then, in Ref.~\cite{Dvo16b}, we applied this approach to study the generation of magnetic fields in a hybrid star (HS) and in a quark star (QS). Moreover, in Ref.~\cite{Dvo16c}, we studied the generation of small-scale magnetic field in quark matter accounting for the electroweak interaction between quarks and the turbulent motion of matter as well as applied the results to describe magnetar bursts. In this work, we summarize the results in Refs.~\cite{DvoSem15a,DvoSem15b,DvoSem15c,Dvo16a,Dvo17,Dvo16b,Dvo16c}.


Let us consider degenerate fermions, which the background matter of a compact star is composed of. We shall assume that the matter density is high enough for
the chiral symmetry to be restored. In this case CME can happen in this system. We showed in Ref.~\cite{Dvo16d} that CME for electrons cannot take place inside a neutron star since the background matter is not sufficiently dense. Nevertheless, the chiral symmetry can be restored in HS/QS~\cite{BubCar16}. In Ref.~\cite{Dvo16b},
accounting for the electroweak interaction between quarks, we showed
that the anomalous electric current $\mathbf{J}_{5}$ along the external
magnetic field $\mathbf{B}$ is induced in this matter,
\begin{equation}\label{eq:Jind}
  \mathbf{J}_{5} = \Pi\mathbf{B},
  \quad
  \Pi = \frac{1}{2\pi^{2}}\sum_{q}e_{q}^{2}
  \left(
    \mu_{5q}+V_{5q}
  \right),
\end{equation}
where $e_{q}$ is the electric charge of a 
quarks, $\mu_{5q}=\left(\mu_{q\mathrm{R}}-\mu_{q\mathrm{L}}\right)/2$
is the chiral imbalance, $\mu_{q\mathrm{R,L}}$ are the chemical potentials
of right and left quarks, $V_{5q}=\left(V_{q\mathrm{L}}-V_{q\mathrm{R}}\right)/2$,
and $V_{q\mathrm{L,R}}$ are the effective potentials of the electroweak
interaction of left and right quarks with background fermions~\cite{Dvo16b,Dvo16c}.

The evolution of the magnetic field is described by the spectra of the magnetic helicity density $h(k,t)$ and the magnetic energy density $\rho_{\mathrm{B}}(k,t)$. The total system of kinetic equations has the form~\cite{Dvo16b,Dvo16c},
\begin{align}\label{eq:syskindm}
  \frac{\partial h(k,t)}{\partial t} = & -2k^{2}
  \left[
    \frac{F_{Q}^{5/6}}{\sigma_{\mathrm{cond}}} +
  \frac{4}{3}\frac{\tau_{\mathrm{D}}}{\rho_{\mathrm{E}}+p}
  \int\mathrm{d}k'\rho_{\mathrm{B}}(k',t)\right]h(k,t)
  \nonumber
  \displaybreak[1]
  \\
  & +
  4
  \bigg[
    \frac{2\alpha_{\mathrm{em}}F_{Q}^{5/6}}{\pi\sigma_{\mathrm{cond}}}
    \left\{
    \frac{4}{9}
      \left[
        \mu_{5u}(t)+V_{5u}
      \right] +
      \frac{1}{9}
      \left[
        \mu_{5d}(t)+V_{5d}
      \right]
    \right\}
    \notag
    \displaybreak[1]
    \\
    & +
    \frac{2}{3}\frac{\tau_{\mathrm{D}}}{\rho_{\mathrm{E}}+p}
    \int\mathrm{d}k' k^{\prime2}h(k',t)
  \bigg]
  \rho_{\mathrm{B}}(k,t),
  \nonumber
  \displaybreak[1]
  \\
  \frac{\partial\rho_{\mathrm{B}}(k,t)}{\partial t} = & 
  -2k^{2}
  \left[
    \frac{F_{Q}^{5/6}}{\sigma_{\mathrm{cond}}} +
    \frac{4}{3}\frac{\tau_{\mathrm{D}}}{\rho_{\mathrm{E}}+p}
    \int\mathrm{d}k'\rho_{\mathrm{B}}(k',t)
  \right]
  \rho_{\mathrm{B}}(k,t)
  \nonumber
  \displaybreak[1]
  \\
  & +
  \bigg[
    \frac{2\alpha_{\mathrm{em}}F_{Q}^{5/6}}{\pi\sigma_{\mathrm{cond}}}
    \left\{
      \frac{4}{9}
      \left[
        \mu_{5u}(t)+V_{5u}
      \right] +
      \frac{1}{9}
      \left[
        \mu_{5d}(t)+V_{5d}
      \right]
    \right\}
    \notag 
    \displaybreak[1]
    \\
    & -
    \frac{2}{3}\frac{\tau_{\mathrm{D}}}{\rho_{\mathrm{E}}+p}
    \int\mathrm{d}k'k^{\prime2}h(k',t)
  \bigg]
  k^{2}h(k,t),
  \nonumber
  \displaybreak[1]
  \\
  \frac{\mathrm{d}\mu_{5q}(t)}{\mathrm{d}t} = &   
  \frac{e_q^2 F_{Q}^{5/6}}
  {4\mu_{q}^{2}\sigma_{\mathrm{cond}}}
  \int\mathrm{d}k
  \bigg[
    k^{2}h(k,t)
    \nonumber
    \displaybreak[1]
    \\
    & -
    \frac{4\alpha_{\mathrm{em}}}{\pi}
    \left\{
      \frac{4}{9}
      \left[
        \mu_{5u}(t)+V_{5u}
      \right] +
      \frac{1}{9}
      \left[
        \mu_{5d}(t)+V_{5d}
      \right]
    \right\}
    \rho_{\mathrm{B}}(k,t)
  \bigg] -
  \Gamma_{q}\mu_{5q}(t),
\end{align}
where $\sigma_{\mathrm{cond}}$ is the conductivity of quark matter, $F_Q$ is the quenching factor~\cite{DvoSem15c,Dvo16a,Dvo17,Dvo16b} accounting for the anticorrelation of the magnetic field and the matter temperature, $\rho_{\mathrm{E}}$ is the energy density of the background matter, $p$ is the matter pressure, and $\Gamma_q$ is the helicity flip rate of quarks in their mutual collisions. To account for the evolution of small-scale magnetic fields, in Eq.~\eqref{eq:syskindm}, we take into account the effects of turbulence governed by the phenomenological drag time parameter $\tau_{\mathrm{D}}$~\cite{Dvo16d}. We shall suppose that the initial spectrum of the magnetic energy density has the Kolmogorov form, $\rho_{\mathrm{B}}(k,t_0) \sim k^{-5/3}$. The initial values of the chiral imbalances are taken as $\mu_{5q}(t_0) = 1\,\text{MeV}$.

\begin{figure}
  \centering
  \subfigure[]
  {\label{1a}
  \includegraphics[scale=.23]{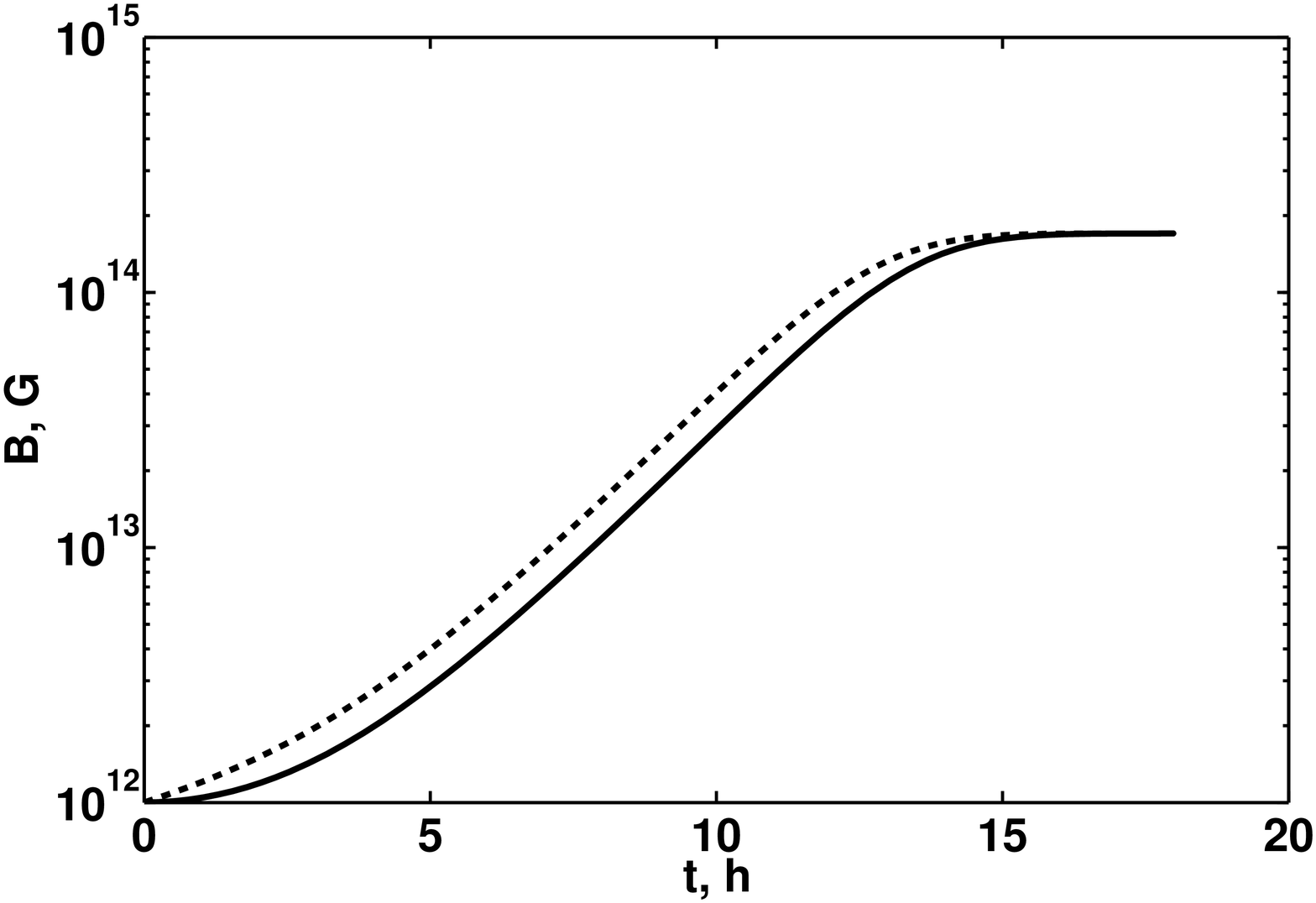}}
  \hskip-.7cm
  \subfigure[]
  {\label{1b}
  \includegraphics[scale=.23]{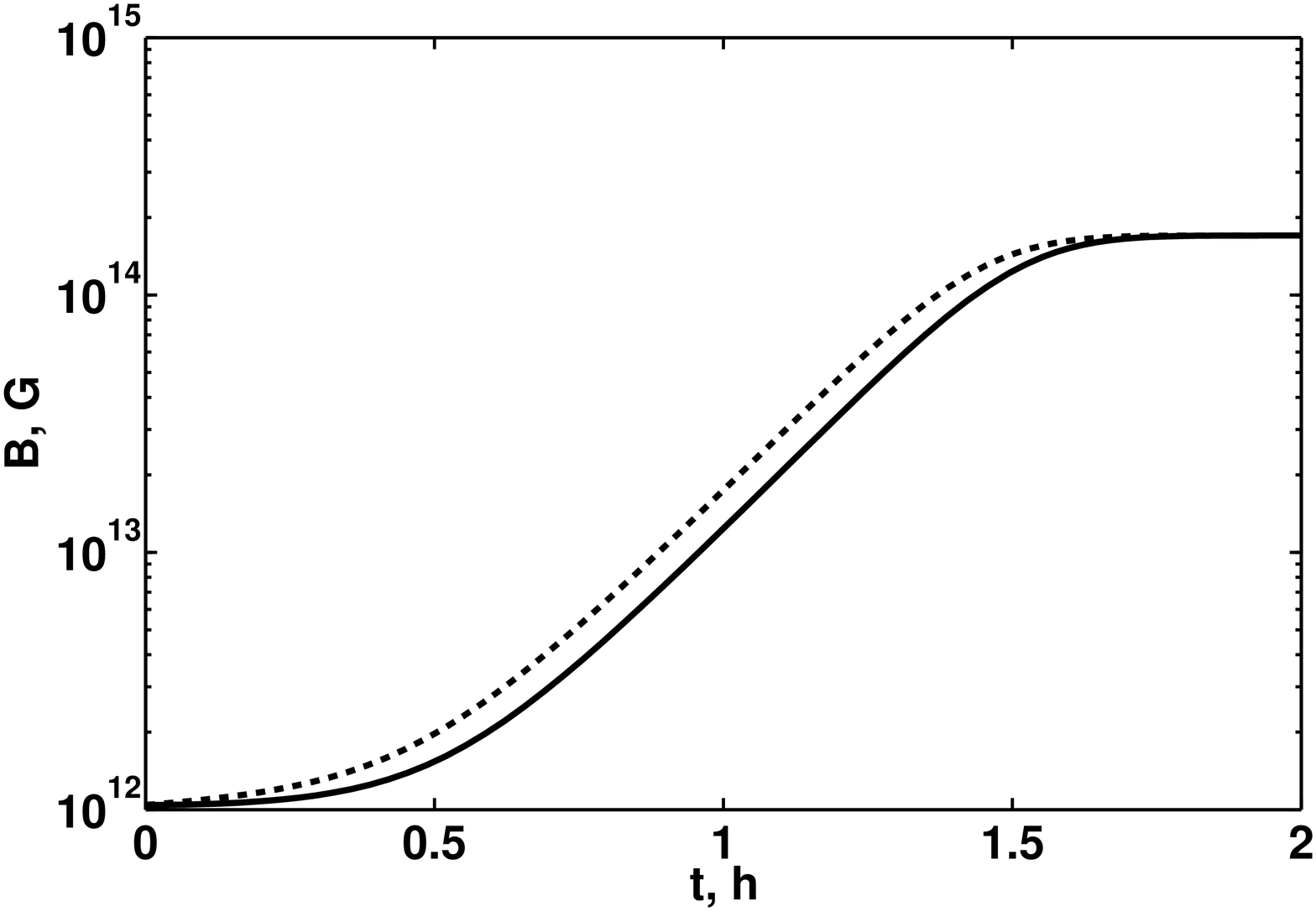}}
  \\
  \subfigure[]
  {\label{1c}
  \includegraphics[scale=.23]{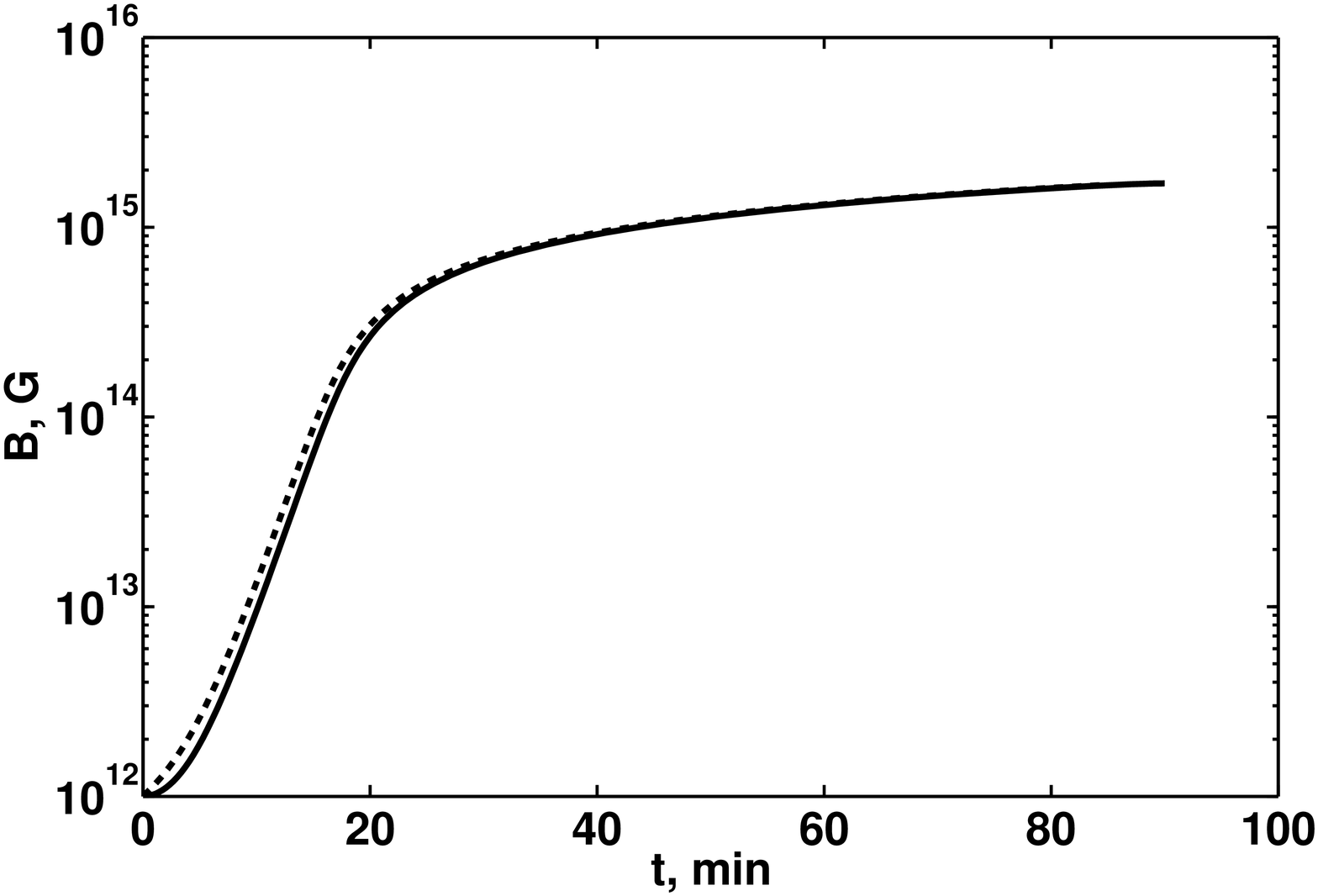}}
  \hskip-.7cm
  \subfigure[]
  {\label{1d}
  \includegraphics[scale=.23]{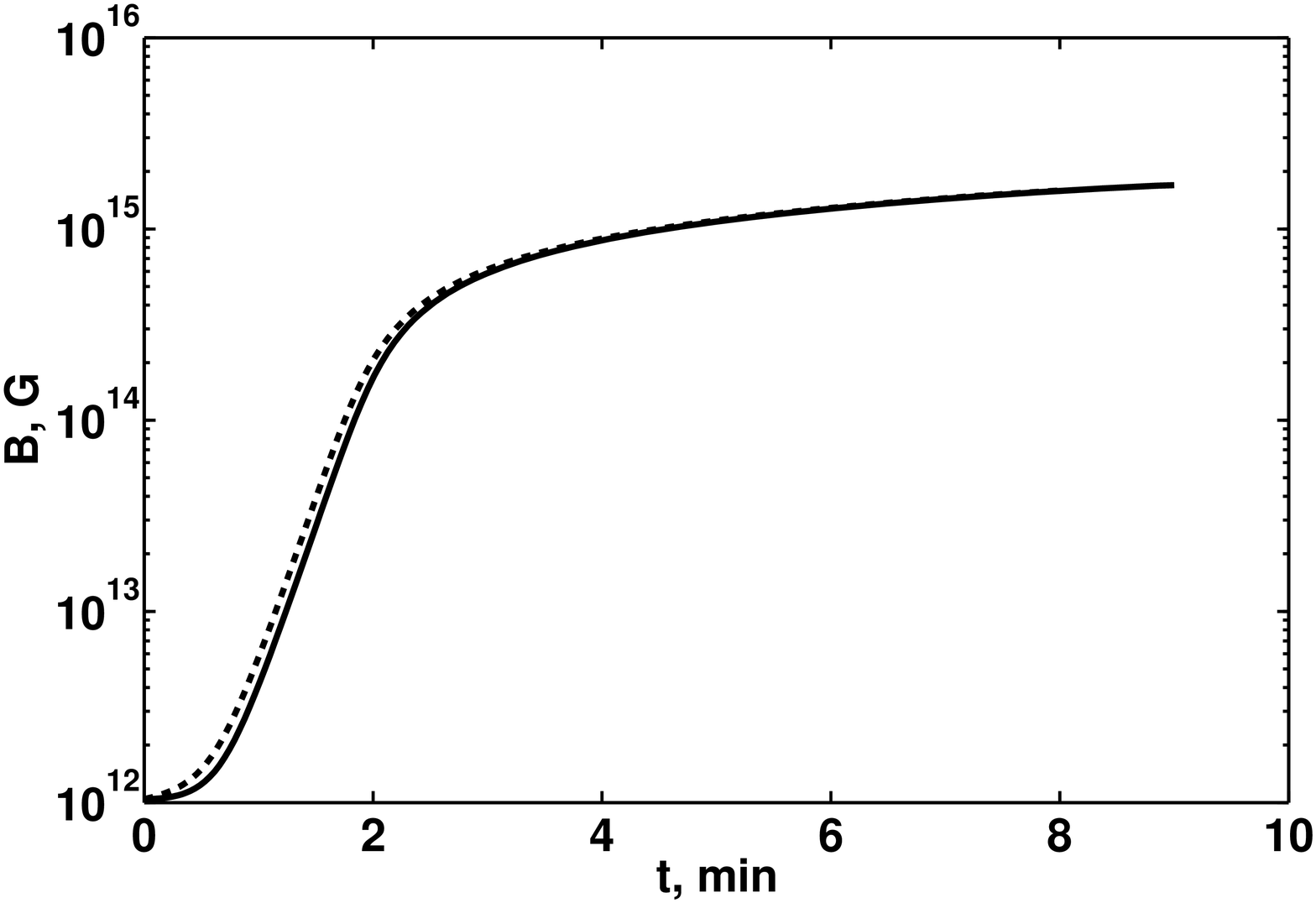}}
  \protect
  \caption{\label{fig:Bevol}
    The magnetic field versus time for different initial temperatures
    $T_{0}$ and $\Lambda_\mathrm{min}$ in quark matter consisting
    of $u$ and $d$ quarks.
    The solid lines correspond to initially nonhelical fields
    and dashed ones to the fields having maximal initial helicity.
    (a)~$T_{0}=10^{8}\,\text{K}$ and   
    $\Lambda_\mathrm{min}=1\,\text{km}$.
    (b)~$T_{0}=10^{8}\,\text{K}$ and   
    $\Lambda_\mathrm{min}=100\,\text{m}$.
    (c)~$T_{0}=10^{9}\,\text{K}$ and 
    $\Lambda_\mathrm{min}=1\,\text{km}$.
    (d)~$T_{0}=10^{9}\,\text{K}$ and 
    $\Lambda_\mathrm{min}=100\,\text{m}$. 
  }
\end{figure}

We present below the numerical solution of Eq.~\eqref{eq:syskindm} in the case of large-scale magnetic field, i.e. we set $\tau_{\mathrm{D}} = 0$.  In this case, the wave number $k$ in Eq.~\eqref{eq:syskindm} changes in the range: $R^{-1} < k < \Lambda_\mathrm{min}^{-1}$, where $R = 10\,\text{km}$ is the compact star radius and $\Lambda_\mathrm{min}$ is a free parameter. In Fig.~\eqref{fig:Bevol}, one can see the time evolution of magnetic fields for different initial temperatures, initial helicities, and various $\Lambda_\mathrm{min}$.

As one can see in Fig.~\ref{fig:Bevol}, the seed magnetic field
$B_{0}=10^{12}\,\text{G}$, which is typical in
a young pulsar, is amplified in quark matter up to $B\sim\left(10^{14}-10^{15}\right)\,\text{G}$,
depending on the initial temperature. It should be noted that we predict the generation of large-scale magnetic field with $10^2\,\text{m} < \Lambda < 10\,\text{km}$.
Such magnetic fields can be found
in magnetars~\cite{Mer15}. Therefore one can conclude that HS/QS can potentially 
become a magnetar.

\begin{figure}
  \centering
  \subfigure[]
  {\label{2a}
  \includegraphics[scale=.11]{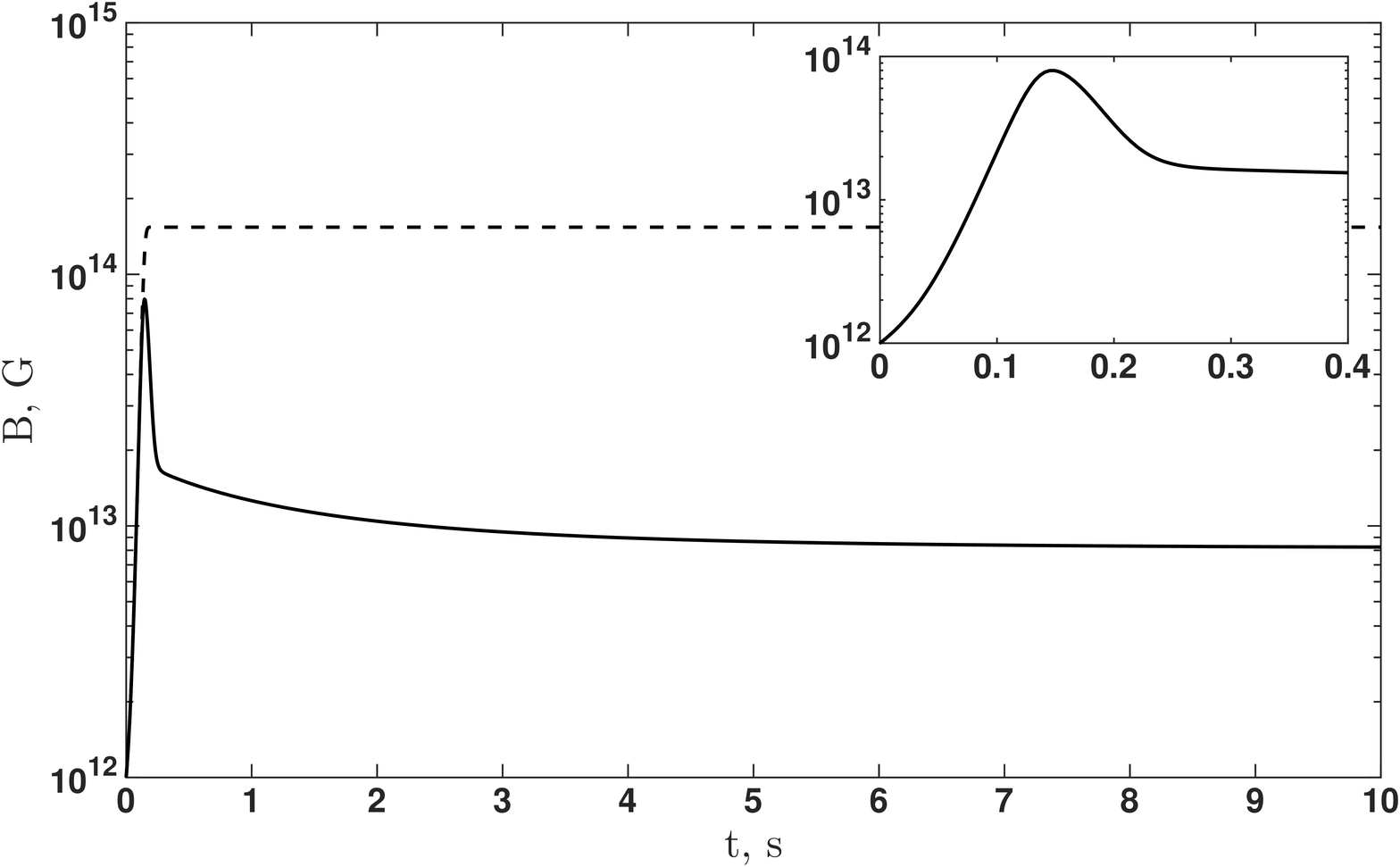}}
  \hskip-.7cm
  \subfigure[]
  {\label{2b}
  \includegraphics[scale=.11]{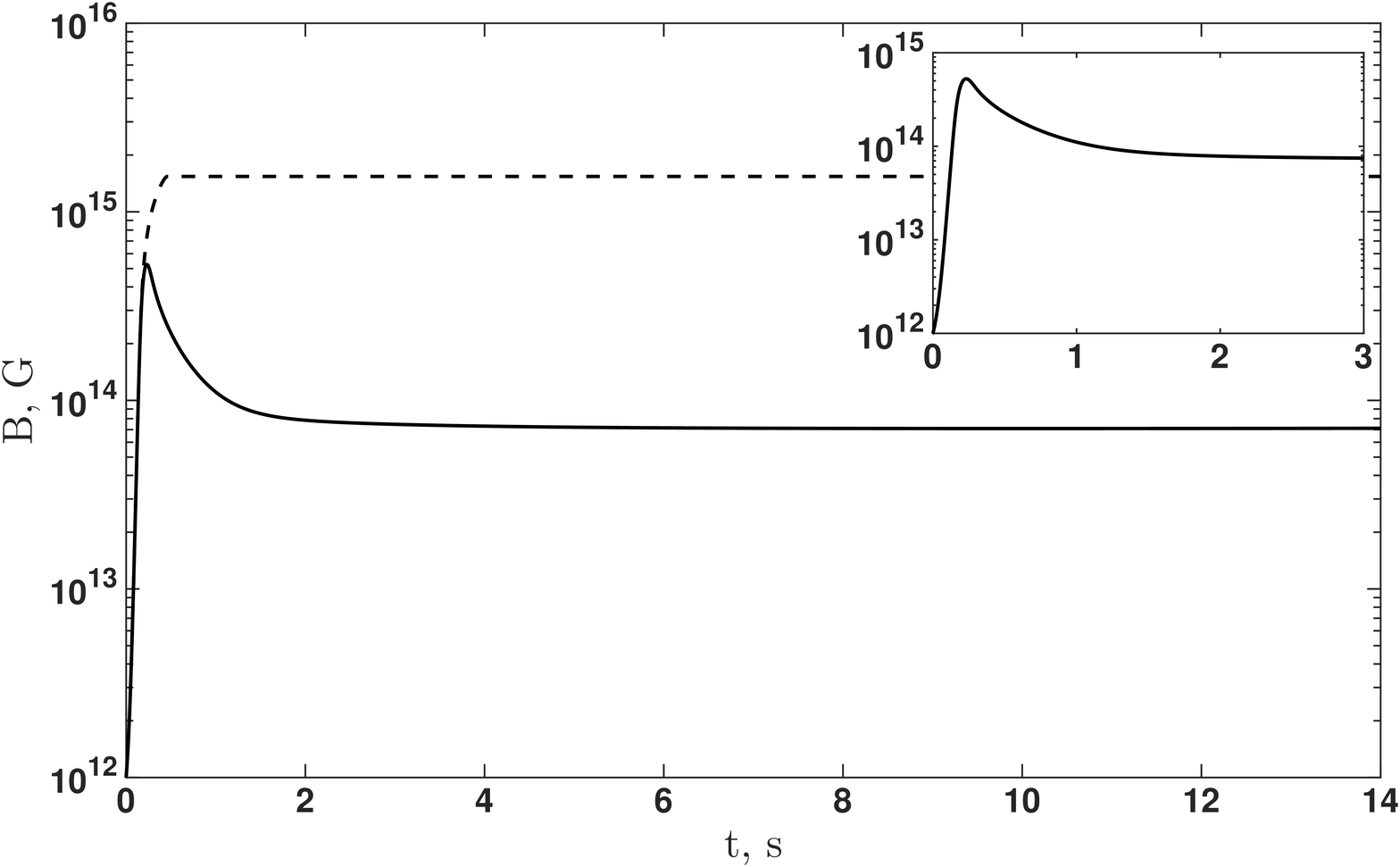}}
  \\
  \subfigure[]
  {\label{2c}
  \includegraphics[scale=.11]{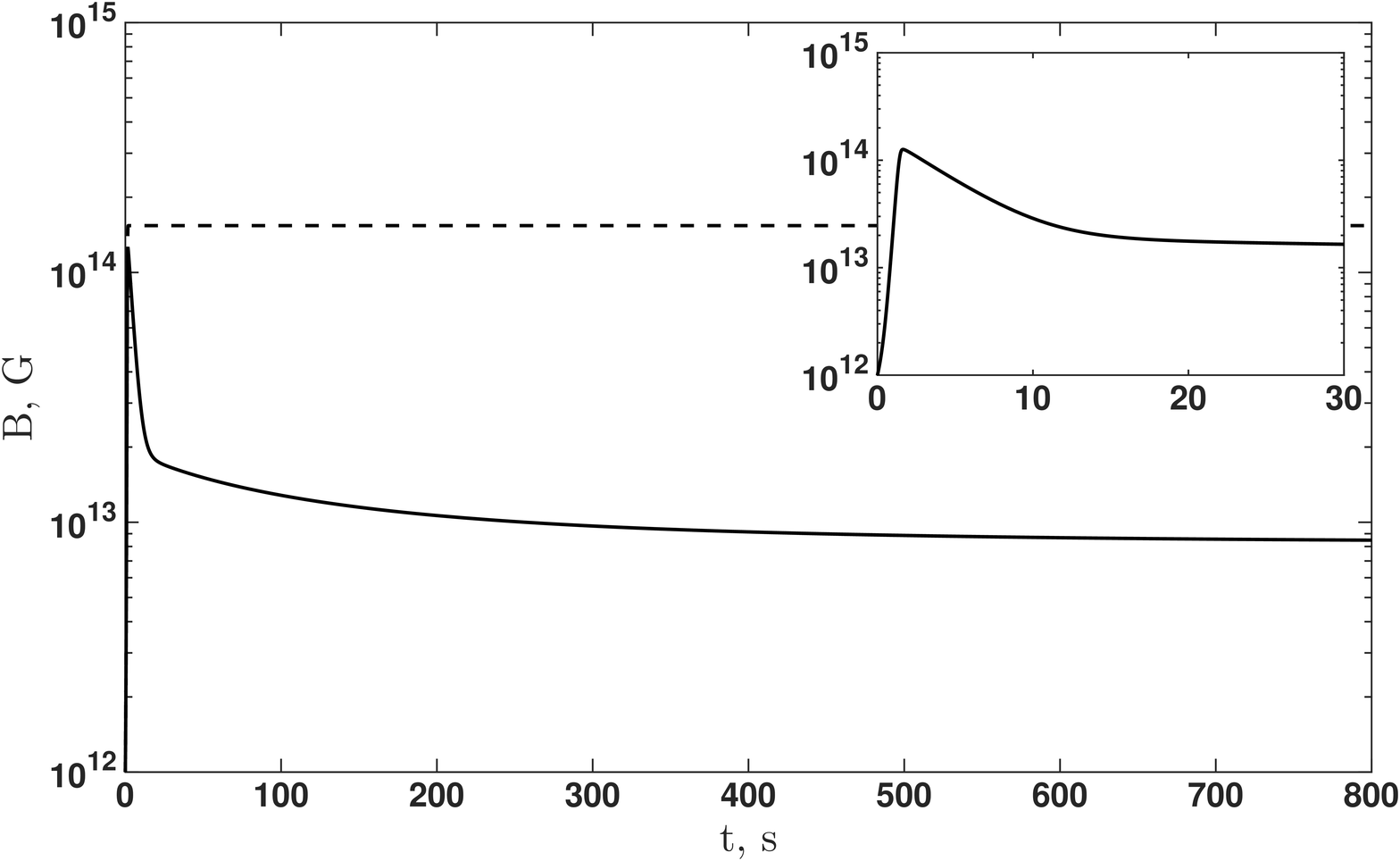}}
  \hskip-.7cm
  \subfigure[]
  {\label{2d}
  \includegraphics[scale=.11]{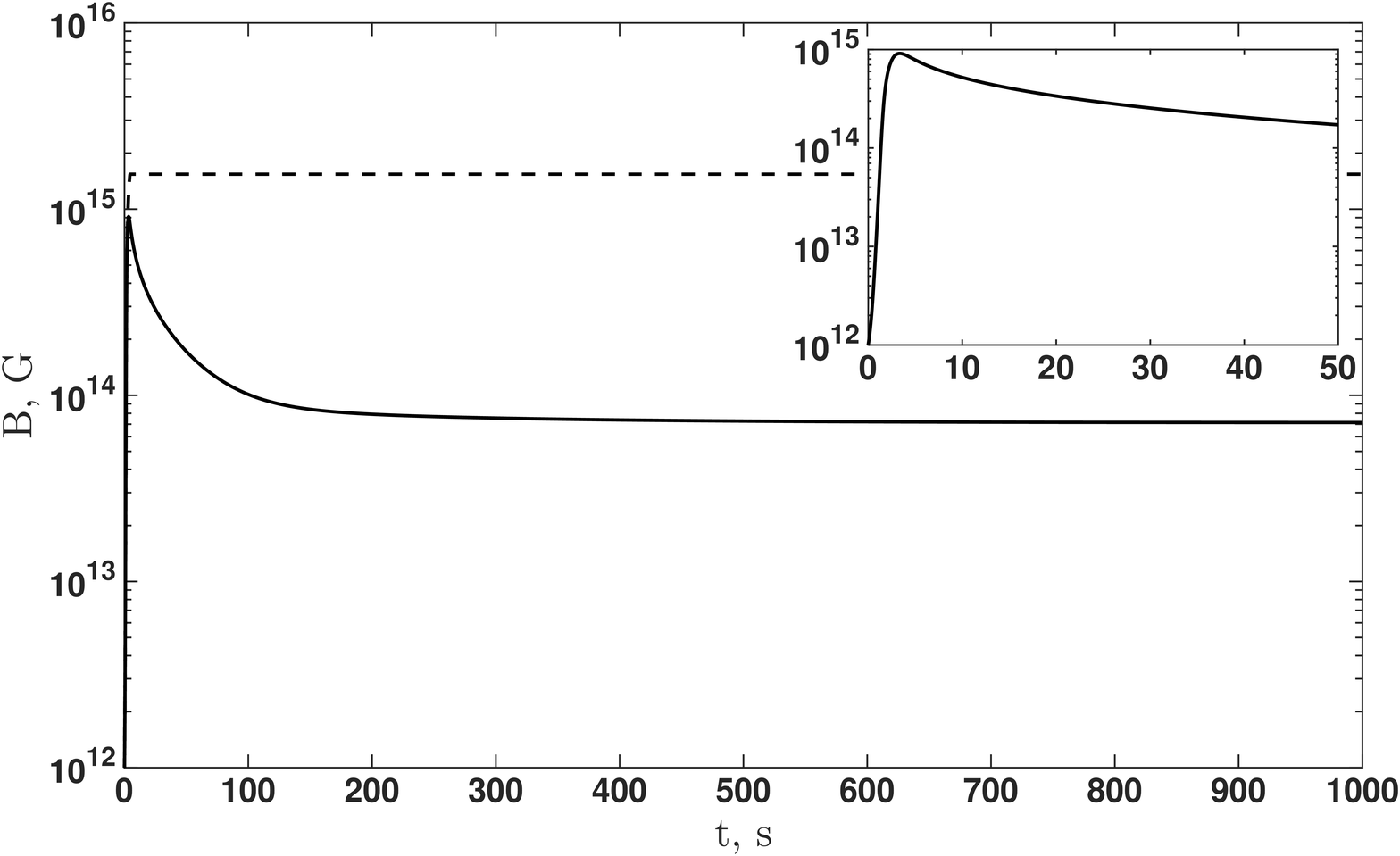}}
  \protect
  \caption{The evolution of the magnetic field for different $T_{0}$ and 
  $\Lambda_{\mathrm{max}}$
  in case when $u$ and $d$ quarks are present in a compact star.
  The panels~(a) and (c) correspond to $T_{0}=10^{8}\,\text{K}$,
  whereas the panels~(b) and (d) to $T_{0}=10^{9}\,\text{K}$.
  The panels~(a) and~(b) show the evolution of the field with
  $1\,\text{cm}<\Lambda<10\,\text{cm}$, whereas
  the panels~(c) and~(d) with
  $10\,\text{cm}<\Lambda<10^{2}\,\text{cm}$.
  The insets demonstrate the magnetic field behavior
  at small evolution times.
  Dashed lines represent the magnetic field evolution with the corresponding 
  initial conditions at the absence of the turbulence.
  \label{fig:Bfield}}
\end{figure}

Now let us turn to the evolution of small-scale magnetic fields. In this situation, we shall take that $\Lambda_\mathrm{min} = \Lambda_\mathrm{max}/10$ and suppose that the initial magnetic field is maximally helical. The time evolution of magnetic fields in turbulent quark matter is shown in Fig.~\ref{fig:Bfield}. One can see that the matter turbulence becomes important as soon as the small-scale magnetic field reaches certain strength. Then the magnetic field starts to decrease compared to the case of non-turbulent matter, shown as dashed line in Fig.~\ref{fig:Bfield}.

The time evolution of small-scale magnetic fields driven by the electroweak interaction of quarks in turbulent quark matter resembles the profiles of electromagnetic flashes of magnetars known as short bursts and giant flashes of AXP and SGR~\cite{Mer15}. These flashes are believed to originate from the twists of magnetic field lines in the magnetar magnetosphere. Such a twist should be related to the plastic deformation of the stellar crust which can be driven by a thermoplastic wave~\cite{BelLev14}. Despite the plausibility of the thermoplastic wave model, the origin of an effect, which triggers the propagation of this wave from the core to the stellar surface, is unclear. We suggest that a fluctuation of a magnetic field near the magnetar core, where the chiral symmetry can be restored, can initiate the propagation of a thermoplastic wave.

In conclusion we mention that,  in Refs.~\cite{DvoSem15a,DvoSem15b,DvoSem15c,Dvo16a,Dvo17}, we developed the model for the generation of strong large-scale magnetic fields in dense matter of a compact star driven by the electroweak interaction between background fermions. Then, in Ref.~\cite{Dvo16b}, this model was applied to generate magnetic fields in dense quark matter in HS/QS. The electromagnetic flashes of magnetars were explained in frames of our model in Ref.~\cite{Dvo16c}.

\end{document}